\documentclass[aip,jap,preprint]{revtex4-1}
\usepackage{epstopdf}
\usepackage{graphicx}  % needed for figures
\usepackage{dcolumn}   % needed for some tables
\usepackage{bm}        % for math
\usepackage{amssymb}
\usepackage[english]{babel}
\usepackage{lineno}

\begin{document}

\title{High pressure anomalies in exfoliated $MoSe_2$: Resonance Raman and X-ray diffraction studies }

\author{Pinku Saha, Bishnupada Ghosh, Aritra Mazumder, and Goutam Dev Mukherjee}

%\author{Bishnupada Ghosh}

%\author{Aritra Mazumder}

%\author{Goutam Dev Mukherjee}
\email [Corresponding Author:]{ goutamdev@iiserkol.ac.in}
\affiliation {National Centre for High Pressure Studies, Department of Physical Sciences, Indian Institute of Science Education and Research Kolkata, Mohanpur Campus, Mohanpur – 741246, Nadia, West Bengal, India.}

%\date{\today}

\begin{abstract}

Detailed high pressure Resonance Raman ($RR$) Spectroscopy and X-ray diffraction ($XRD$) studies are carried out on 3-4 layered $MoSe_2$ obtained by liquid exfoliation. Analysis of ambient $XRD$ pattern and $RR$ spectra indicate the presence of a triclinic phase along with its parent hexagonal phase. Pressure evolution of prominent Raman modes and their full width at half maximum ($FWHM$) show slope changes at about 13 GPa and 33 GPa, respectively. Slope change in the linear behavior of reduced pressure ($H$) with respect to Eulerian strain ($f_E$) is observed at about 13 GPa. A minimum in the $FWHM$ values of $E_{2g}^1$ and $A_{2u}^2$ modes at the same pressure indicate to an electronic topological transition ($ETT$). Above 33 GPa the sample completely gets converted to the triclinic structure, which indicates the importance of strain in structural as well as electronic properties of two dimensional materials.

\noindent {\bf Keywords:} {Transition metal dichalcogenides, high pressure, Raman Spectroscopy, X-ray diffraction, crystal structure }

\end{abstract}

\maketitle

\section{Introduction}
Layered wide bandgap semiconductors have attracted interest as their electronic and optical properties can be modified by bandgap engineering. In recent years strain engineering or introduction of defects have been found to modify the band structure of these 2D materials leading to novel electronic properties. Introduction of strain has been found to induce electronic structural transition and enhance the electronic properties in $TiS_2$ and graphene\cite{samanta14,huang10}. Tuning of optical band gap with strain in layered two-dimensional (2D) $InSe$ has shown it to be a versatile optoelectronic material\cite{song18}. In a recent strain engineering work on the $MoSe_2$/$WSe_2$ heterolayers, it has been shown that decreasing the layer distance by application of uniaxial pressure can change the conductance by several orders of magnitude\cite{sharma14}. In a detailed study on layered $MoS_2$ transition-metal-dichalcogenide (TMD) Liu {\it et al.}\cite{liu14} have mapped the strain and structural heterogeneity with shift of Photoluminescence (PL) peak and have shown the importance of strain induced bandgap engineering. A theoretical simulation study based on state-of-the-art density functional calculations including vibrational energy corrections have shown relatively small amount of tensile strain under uniaxial conditions can induce semiconducting to metallic transition in $Mo-$ and $W-$ dichalcogenide compounds\cite{duerloo14}.  

Pressure can be an excellent tool to study the structural and electronic phase transitions in layered TMD's to understand their behaviour under strained conditions. There are several reports suggesting metallization of TMD's under pressure\cite{nayak14,nayak2014,chi14,hromadova13,zhao15,riflikova14,bandaru14,duwal16,nayak15,saha18}. Some of the above reports suggest possible structural transition along with metallization with pressure.
High pressure Raman, electrical resistance, and synchrotron x-ray diffraction ($XRD$) studies on hexagonal 2$H-MoSe_2$ (single crystal and powder) do not report any evidence of structural transition accompanying the electronic phase transition\cite{sugai82,dave04,aksoy08}. Later a systematic high pressure investigation on single crystal $MoSe_2$ up to 60 GPa using multiple experimental techniques and {\it{ab-initio}} calculations reported pressure induced metallization above 40 GPa, but no structural transition\cite{zhao15}. However, Aksoy {\it et al.}\cite{aksoy08} have observed a discontinuity in the ratio of $(\Delta c/c_0)$/$(\Delta a/a_0)$ ($c$ and $a$ are the lattice parameters with subscript 0 referring to ambient pressure) at 10 GPa.  A recent theoretical study predict tetragonal phase with space group $P4/mmm$ to become stable in $MoSe_2$ above 118 GPa\cite{kohulak17}.    
Though the above studies on layered $MoSe_2$ show metallization of the sample without any structural transition under stress, a complete experimental characterization of the exfoliated $MoSe_2$ sample at high pressure is lacking. Exfoliated $MoSe_2$ layers can develop inherent surface strain similar to $MoS_2$ and $WS_2$\cite{pal17,saha18} and therefore a careful detailed high pressure study is needed to address the effect of strain in the structural and electronic  behavior of exfoliated $MoSe_2$.

In the present work, we have carried out a detailed high pressure investigation on exfoliated layered $MoSe_2$ using Raman spectroscopy and $XRD$ measurements up to about 50 GPa. Anomalies in the positions and full width at half maximum ($FWHM$) of Raman modes are observed at about 13 GPa and 33 GPa, respectively. $XRD$ studies in conjunction with Raman anomalies indicate to the presence of electronic topological transition at about 13 GPa followed by a phase transformation from hexagonal structure to a triclinic crystal structure at high pressures, similar to that observed in exfoliated $WS_2$.

\section {Experimental}
 Few layers of $MoSe_2$ sample are prepared by liquid exfoliation\cite{saha18,mayoral16} of crystalline $MoSe_2$ (from Sigma Aldrich, powder, -325 mesh, 99.9\% trace metals basis). After exfoliation samples of approximate size 10-15 $\mu$m are chosen for loading in the $DAC$. We have used a piston-cylinder type $DAC$ from EasyLab Co. (UK) for our high pressure studies. Sample is loaded inside the central hole of diameter 100 $\mu$m of stainless gasket preindented to 45 $\mu$m. Mixture of methanol-ethanol at a ratio of 4:1 is used as pressure transmitting medium for high pressure Raman measurements. For the determination of pressure\cite{mao86} few ruby chips (approximate size 3-5 micron) are loaded along with the sample. Raman spectra are collected in the back scattering geometry using a micro-Raman system (LabRam HR 800) with 1800 g/mm grating and a spectral resolution better than 1.2 cm$^{-1}$. Appropriate edge filters are used for Raman measurements from 50 $cm^{-1}$, which are provided by Horiba Jobin Yvon. Ambient pressure Raman measurements are carried out at four different excitation wavelengths, 488, 532,  633 and 785 nm using a 100X microscope objective. High pressure Raman measurements are carried out at 488 nm excitation energy using 20X long working distance objective (infinitely corrected). The laser power is kept at 16 mW at ambient pressure and then increased to 40 mW at high pressure to obtain a good signal to noise ratio and also to avoid local heating of the sample.

For high pressure $XRD$ measurements, the exfoliated samples are loaded in the $DAC$ using the same procedure described above. In this case, small amount  of silver powder is loaded along with sample for pressure estimation\cite{dewaele08}. $XRD$ studies are carried out at room temperature at the XPRESS beamline in Elettra synchrotron source, Italy using monochromatic X-ray radiation of wavelength 0.4957 \AA. The X-ray beam is collimated to 20 micron and the diffracted X-rays are detected using a MAR 3450 image plate type detector aligned normal to the beam. The sample to detector distance is calibrated using $LaB_6$. The two dimensional diffraction images are integrated to get intensity versus 2$\theta$ profile using Dioptas software\cite{prescher15}. The XRD data are then indexed using the CRYSFIRE\cite{shirley02} and CHECKCELL program \cite{checkcell} followed by LeBail fitting using GSAS\cite{toby01}.

\section{Results and Discussion}

Initially the exfoliated sample is placed on the diamond culet and Raman spectra are collected  to have an idea of resonance condition and layer thickness. Raman spectra of the exfoliated sample at ambient condition and different excitation energies are shown in Fig.1(a). In the right-inset of Fig.1(a) we have shown the magnified image of the exfoliated sample of approximate size 10-15 $\mu$m on the diamond anvil culet of size 300 $\mu$m.
2$H-MoSe_2$ has a space group symmetry of $D_{6h}^4$ and there are 12 modes of lattice vibrations at the centre of Brillouin zone\cite{sekine80,nam15}: $A_{1g} + 2A_{2u} + B_{1u} +2B_{2g} + E_{1g} + 2E_{1u} + 2E_{2g} + E_{2u}$. Among these $A_{1g}$, $E_{1g}$, $E_{2g}^1$, and $E_{2g}^2$ are Raman active modes \cite{sugai82}. At 785 and 633 nm laser excitations we find a prominent mode at 242 $cm^{-1}$  along with another broad mode at 147 $cm^{-1}$. At laser excitations of 532 and 488 nm, in addition to the above, few other prominent modes are seen at 169, 285, 352 $cm^{-1}$. Several other additional Raman modes are seen in case of 488 nm laser excitation (Fig.1(b)). The mode at 147  $cm^{-1}$ can be seen in case of 532 nm laser excitation, however is not much prominent at 488 nm laser excitation due to several other sharp Raman modes below 200 $cm^{-1}$. The modes at  242 and 285 $cm^{-1}$ are identified as two first order Raman modes  $A_{1g}$ and $E_{2g}^1$, respectively \cite{zhao15,sugai82}. The mode at 147 $cm^{-1}$ is assigned as $E_{2g}^1 - LA$, which is close to Stokes-anti-Stokes pair at the M point as indicated by Nam {\it et al.} \cite{nam15}.  The laser excitation at 488 nm corresponds to the energy 2.54 eV, which is very close to the band-to-band transition at 2.5 eV \cite{sekine80,nam15,soubelet16,sugai82}. Therefore, additional modes are excited due to the resonance condition, which are in agreement with the other measurements on $MoSe_2$\cite{sekine80,nam15,larentis12}.
Intensity of $A_{1g}$ mode is found to be enhanced with respect to that of $E_{2g}^1$ in agreement with the literature\cite{sekine80,nam15}. Even though the $E_{1g}$ mode is forbidden in back scattering geometry, it is found to be prominent around 169 $cm^{-1}$. Similarly the mode observed at 352 $cm^{-1}$ can be indexed to the infrared active mode, $A_{2u}^2$. Above anomalous observation of modes can be related to resonance condition\cite{sekine80,nam15,soubelet16}.  
 Similar phenomena are also observed in nitrogen doped GaAs\cite{cheong00}. A close inspection reveals a shoulder on the left side of $A_{1g}$ mode, which can be identified as a Raman mode centered at 239 $cm^{-1}$ (see  right-inset of Fig.1). This mode is  labeled as $B_{1u}$\cite{sekine80,nam15}. Splitting of $A_{1g}$ mode has been shown to depend on the layer thickness \cite{soubelet16}. In the present case double splitting of $A_{1g}$ mode indicates to three to four layers of $MoSe_2$ in our exfoliated sample \cite{soubelet16}. 
Few additional modes in the range from 75 to 140 $cm^{-1}$ and humps around 215 and 360 $cm^{-1}$ are observed in the Raman spectrum (indicated inside dotted red rectangle in Fig.1). The prominent additional modes below $E_{1g}$mode are found at 91, 94, 99, 107, 116, 122, 130, 138, 147, 150 and 154 $cm^{-1}$. Similar features are also found in 1-5 layers of $MoSe_2$ using excitation wavelength 465 nm by Soubelet {\it et al.}\cite{soubelet16} but remain unresolved.  
Additional Raman modes observed in earlier Raman studies on exfoliated $WS_2$ and $MoS_2$ are assigned to different crystal structures embedded in parent hexagonal phase \cite{saha18,pal17}. Therefore it is important to discuss XRD measurements before proceeding to further analysis of high pressure measurements.

In Fig.2 we have shown the ambient pressure XRD pattern of the exfoliated sample and compared with XRD patterns collected at high pressures. All the intense reflection lines at ambient condition are indexed to a hexagonal structure with space group $P6_3/mmc$ and lattice parameters $a$ = 3.288(8) \AA, $c$ = 12.958(6) \AA  with volume = 121.36(6) $\AA^3$ and $Z$ = 2, which is consistent with the literature\cite{aksoy08,zhao15}. In addition a few new peaks are observed in the ambient pattern, which are indicated by the black arrow in the Fig.2 and could only be indexed to a triclinic cell. The triclinic phase is indexed to space group $P\overline{1}$ with lattice parameters: $a$ =6.991(4), $b$ = 7.151(6), $c$ = 7.054(5) \AA, $\alpha$ = 126.1(2)$^\circ$, $\beta$ = 88.4(2)$^\circ$, $\gamma$ = 105.3(2)$^\circ$, Volume = 271.04(7) \AA$^3$ and $Z$ = 4 and it is consistent with the literature\cite{mahler14,saha18}. In Fig.3(a) we have shown Le-Bail refinement of the triclinic phase along with the Rietveld fit of the hexagonal phase. Le-Bail refinement of the triclinic phase is carried out in the absence of a structural model. The fitting of XRD pattern indicates a mixed phase for exfoliated sample under ambient condition it-self with a small amount of triclinic phase along with the parent hexagonal phase.

Pressure evolution of XRD patterns (Fig.2) show that the intensity of all the Bragg peaks of our sample rapidly decrease with pressure above 12.3 GPa and start broadening extensively. A close inspection reveals that the Bragg-peaks corresponding to triclinic phase become more evident in expense of Bragg-peaks corresponding to hexagonal phase with increase in pressure. Above 33 GPa the XRD patterns could only be indexed to triclinic crystal structure and the corresponding profile fit along with $XRD$ pattern is given in Fig.3(b). This represents a transition from a mixed phase to a pure phase. This result is in contrast to earlier high pressure $XRD$ studies which did not report any structural transition in $MoSe_2$\cite{aksoy08,zhao15}. However a close inspection of data reported by Aksoy et.al. \cite{aksoy08} shows presence of a Bragg peak close to (103) line (similar to our results), which however was ignored in the analysis. 

In Fig.4(a) we have shown the pressure evolution of $c/a$ value of the hexagonal phase. Large value of $c/a$ ratio show that the crystal structure is composed of planes of atoms loosely packed along $c$-axis. Therefore one would expect continuous decrease in $c/a$ ratio with pressure. Instead the data show a step like characteristic with almost constant value in the pressure range 12 - 18 GPa indicating an anisotropic compression below 12 GPa and above 18 GPa. 
The anomalous change in the $c/a$-ratio starts just above the freezing pressure of liquid pressure transmitting medium leading to quasi-hydrostatic environment. Therefore the effect of non-hydrostatic compression on the sample would have lead to either a slope change or a sudden change in the behaviour of $c/a$-ratio instead of the observed step-like behaviour in the pressure range 12-18 GPa. Our XRD measurements are carried out by loading very small amount of sample compared to the total volume of the gasket hole and the XRD patterns are taken from the centre of the diamond anvil cell from an area of diameter 20 micron. Hence effect  of non-hydrostatic component on the sample can be neglected.  Therefore a large change in strain in the lattice is expected due to the anisotropic compression as the volume do not show any anomalous behaviour with pressure (Fig.4(b)). 
 To look into the effect of internal strain we have plotted reduced pressure $H$ $(=\frac{P}{3f_E(1+2f_E)^{5/2}})$ with respect to the Eulerian strain $f_E$ ($=\frac{1}{2}[(\frac{V_0}{V})^{2/3}-1]$)\cite{angel01,poilan11} and shown in Fig.4(c) and (d) for hexagonal and triclinic phases, respectively. $H$ {\it vs} $f_E$ relation is linear  as shown in the following equation considering third order Birch-Murnaghan equation of state\cite{birch47,murnaghan37}:

\begin{equation}
H=K_0 + \frac{3}{2}K_0(K^{'}-4)f_E
\end{equation}
\noindent
where, $V_0$ is the volume at 1 bar and 300 K, $V$ is the volume at pressure $P$, $K_0$ is the bulk modulus, and $K^{'}$ is the first derivative of bulk modulus. However, slope changes  in both the phases are observed around 13 GPa (shown by black arrow in the Fig.2(c) and (d)). Similar anomalies have been observed in different systems during electronic topological transition ($ETT$)\cite{jana16,poilan11,ardit10}. Linear fits to the data up to 13 GPa provide $K_0$ = 37.4(7) GPa and $K^{'}$ = 11.6(4) for hexagonal phase and $K_0$ = 72.5(9) GPa and $K^{'}$ = 13.8(4) for triclinic phase. Above 13 GPa the fits provide $K_0$ = 53.4(5) GPa and $K^{'}$ = 4.6(1) for hexagonal phase and $K_0$ = 100.2(8) GPa and $K^{'}$ = 4.6(1) for triclinic phase. Reported values of Bulk modulus and its pressure derivative for MoSe2 are: (i) as reported by Zhao et.al. \cite{zhao15} are 62(1) GPa, 5.6(1), respectively and (ii) as reported by Aksoy et.al. \cite{aksoy08} are 45.7(3) GPa, 11.6(1), respectively. These values are close to our results of the hexagonal phase. 
Hexagonal phase is found to be more compressible as expected, due to layered structure in comparison with triclinic phase. In fact the Eulerian strain value of hexagonal phase at 13 GPa is more than that of triclinic phase indicating increase in disorder in the hexagonal phase. This can be due to the growth of triclinic phase in expense of hexagonal phase. The pressure derivative of bulk moduli are found to be large in both the phases up to 13 GPa compared to those above 13 GPa. Larger pressure dependence of compressibility at low pressures can lead to disorder in the crystal structure. $K^{'}$ value of both the phases in the $1^{st}$ range of pressure (0-13 GPa) and $K_0$ value of both the phases in the $2^{nd}$ range of pressure (above 13 GPa) are consistent with the study on $WS_2$\cite{saha18}. 

In Fig.5 we have plotted the pressure dependence of certain normalized $d$-spacing values of the sample with respect to the $d_{Ag}$-spacing of (111) lattice plane of Ag pressure marker. Pressure evolution of the normalized $d$-spacings corresponding to the Bragg peaks at $2\Theta$ = 4.378$^o$, 11.973$^o$, and 14.874$^o$ also show slope changes at 13 GPa and 33 GPa. Changes in the slope values indicate to a change in the internal strain of our sample with pressure.

In Fig.6 we have shown the pressure evolution of $RR$ spectra at selected pressure points. All the modes harden with pressure as evident from the figure. The additional modes reduce in intensity and disappear as pressure increases. Pressure induced changes in lattice parameters can modify the band-structure of the sample, which is expected to affect inter-band transitions. Also growth of triclinic phase with pressure  affects the resonance condition. Therefore the low intensity resonance Raman modes observed under ambient condition reduce in intensity at high pressures and are masked by the background.  
Intensities of $E_{1g}$ and $A_{2u}$ modes decrease with pressure, while integrated intensity of $A_{1g}$ mode is found to show maximum at about 12.5 GPa. Since $c/a$-ratio is found to decrease by a large value, it affects the out of plane vibrations that are associated with $A_{1g}$-mode and it becomes stronger till about 12 GPa where the $c/a$ ratio levels off. Since the triclinic phase grows in expense of the hexagonal phase with pressure, the characteristic $A_{1g}$ mode does not increase again above 18 GPa where $c/a$-ratio decreases again.
Above 33 GPa intensity of the $E_{2g}^1$ mode corresponding to the in plane vibrations of Mo and Se atoms is found to increase with respect to $A_{1g}$ mode as evident from the figure. At the same pressure the sample transforms completely to triclinic phase and this behaviour can  be related to the same.
A new mode at 103 cm$^{-1}$, and another one at the right shoulder of $E_{1g}$ mode at about 190 cm$^{-1}$ are found to emerge at about 12.5 GPa. Above 33 GPa one more new mode at 317 $cm^{-1}$ is found to emerge. These modes have not been observed in the previous studies\cite{sugai82,zhao15}. We name these modes as $M_1$, $M_2$, and $M_3$. From XRD studies we have found the existence of a triclinic phase along with parent hexagonal phase, which grows with pressure. Therefore in all possibility these modes belong to the triclinic phase, which are masked at low pressures. They start showing up as the modes related to hexagonal phases start decreasing.

Our linear fit of the prominent $A_{1g}$, $E_{1g}$, $A_{2u}^2$, and $E_{2g}^1$ Raman modes do show three linear regions: (i) ambient to 13 GPa, (ii) 13 to 33 GPa and (iii) 33 GPa and above with slope changes in each region (Fig.7(a) and (b)). First slope change at 13 GPa coincides with the pressure value of observed anomaly in Eulerian strain and the second slope change at 33 GPa coincides with the crystal structure change. Reduced slope values above 33 GPa do indicate lattice stiffening at high pressures. Slopes at three different ranges of pressure for these modes are given in Table-I. Such anomalies in linear pressure behavior of Raman modes are not reported in the previous studies on single crystal $MoSe_2$\cite{zhao15,caramazza17}. However, a close inspection of their data do reveal the deviation from linear behavior. 

Anharmonic lattice vibrations can be understood from the volume dependence of frequency through Gr\"{u}neisen parameter ($\gamma$) values:
\begin{equation}
\gamma = - \frac{\delta ln \omega}{\delta ln V} = K \frac{1}{\omega}\left(\frac{\delta \omega}{\delta P}\right)
\end{equation}
\noindent   
where, $\omega$ is the Raman mode frequency, $V$ is the volume at pressure $P$, $K$ is the bulk modulus, and $\frac{\delta \omega}{\delta P}$ is the slope of the pressure dependent Raman modes. For $A_{1g}$, $E_{1g}$, $A_{2u}^2$, and $E_{2g}^1$ prominent modes the Gr\"{u}neisen parameters values are given in the Table-I. The values of $K$ and $\frac{\delta \omega}{\delta P}$ are taken from our study. It is evident from the Table-I that the Gr\"{u}neisen parameter systematically decreases with pressure. As $\gamma$ is associated with the lattice anharmonicity, it seem to decrease with pressure. This implies increase in bond strength that is reflected in continuous structural change from layered hexagonal to three dimensional triclinic structure.
 
In Fig.8, we have plotted $FWHM$ of prominent $A_{1g}$, $E_{1g}$, $A_{2u}^2$, and $E_{2g}^1$ Raman modes with pressure. $FWHM$ of $A_{1g}$, and $E_{1g}$ modes show slope changes in their linear behavior at about 13 GPa followed by a sudden drop at about 35 GPa (Fig. 8(a)). Similar sudden discontinuity in FWHM of Raman modes are observed in layered $WS_2$, and $MoS_2$ at metallization pressure\cite{nayak2014,saha18}. $FWHM$ of $A_{2u}^2$, and $E_{2g}^1$ modes show an interesting behavior with pressure, showing a minimum at about 13 GPa, followed by a sudden jump at about 33 GPa with stronger pressure dependence. Similar minimum in the $FWHM$ of $E_{2g}^1$ mode has been observed in crystal powder of $MoSe_2$\cite{caramazza17} around 18 GPa.
The FWHM of Raman peaks ($\Gamma$) are related to lifetime of phonon modes ($\tau$) as:
\begin{equation}
\Gamma = \frac{1}{\tau}
\end{equation}
In the absence of any structural transition, pressure induced minimum in Raman mode bandwidths are generally associated with an increase in the life time of phonon-modes due to electronic topological transitions\cite{jana16,bera13}.
The life time of a phonon mode can be related to it's electron phonon coupling($\lambda$) via the FWHM ($\Gamma$) of it's Raman mode via %Cite Black Phosphorus%:
\begin{equation}
\lambda = \frac{\Gamma}{\pi \hbar N(E_f)\omega_i^2}
\end{equation}
\noindent
where $\omega_i$ is the peak position and $N(E_f)$ is the density of states which we can consider to be constant for a 2D material giving us a much simpler relation:
\begin{equation}
\lambda \propto \frac{\Gamma}{\omega_i^2}
\label{e7}
\end{equation}
\noindent
In Fig.8(c) we have shown the change in electron phonon coupling of $A^2_{2u}$ and $E^1_{2g}$ with pressure. We clearly see the electron phonon coupling changes with pressure, showing a dip around the 13 GPa pressure point reflecting similar behaviour of FWHM. Similar dip in electron phonon coupling observed in Black Phosphorus has been attributed to ETT \cite{gupta17}.
Pressure dependent resistivity measurements by Zhao {\it et al.} show a step like characteristic behaviour in the pressure range about 12 - 20 GPa similar to the $c/a$-ratio of the hexagonal phase\cite{zhao15}. Our studies show that there is an ever increasing strain component with pressure (Fig.4). This in combination with the dip in FWHM of Raman modes and electron phonon coupling indicates the presence of an ETT around 13 GPa. Similar anomalies have been observed earlier during ETT in other systems \cite{jana16,bera13,poilan11,ardit10,gupta17} and are probably a precursor to semiconductor to metallic transition. Our study shows the importance of careful structural characterization of the 2D layered systems to understand the effect of lattice strain and its relation to the electronic properties.

\section{Conclusions}

In the present study, we have carried out detailed high pressure Raman and $XRD$ investigations on exfoliated $MoSe_2$. Raman spectroscopy measurements indicate the exfoliated sample to consist of 3 to 4 layers. The exfoliated sample is found to have triclinic phase embedded with its parent hexagonal phase, even at the ambient condition. Change in the slope of prominent Raman modes, anomalies in the FWHM of Raman modes, and the change in the slope in the reduced pressure behavior with Eulerian strain indicate to a possibility  $ETT$ at about 13 GPa. High pressure Raman spectroscopy studies show the emergence of a new modes $M_1$, $M_2$, $M_3$, which along with XRD measurements show a continuous change from hexagonal to triclinic crystal structure driven by internal lattice strain. 

%\end{document}

\begin{acknowledgments}
GDM wishes to thank Ministry of Earth Sciences, Government of India for financial support. PS and BG wish to thank DST, INSPIRE program by Department of Science and Technology, Government of India for financial support. The authors also gratefully acknowledge the financial support from the Department of Science and Technology, Government of India for carrying out experiments in the XPRESS beam line in the ELETTRA Synchrotron light source under Indo-Italian Executive Programme of Scientific and Technological Cooperation.
\end{acknowledgments}

\noindent
{\bf {Author Contributions}} All authors has equal contribution. All authors reviewed the manuscript.

\section{Additional Information}

{\bf {Competing Financial Interests:}} The authors declare no competing financial interests.

%%\newpage

\newpage
\begin{table}
\caption{\label{tab:table1} Slope of different modes and the Gr\"{u}neisen parameters ($\gamma_i$) at three different ranges of pressure.}
\begin{ruledtabular}
\begin{tabular}{c|cccc}
& Modes & 0-13 GPa &13-32 GPa&32-49 GPa \\
\hline
&$E_{1g}$& 1.7& 1.0& 0.7 \\
Slope& $A_{1g}$& 2.6& 1.6 & 0.9   \\
($cm^{-1}$/GPa)& $E_{2g}^1$ & 1.9& 0.9 & 0.7 \\
& $A_{2u}^2$ & 2.4& 1.4 & 0.9 \\
\hline
&$E_{1g}$& 0.729& 0.593& 0.415 \\
$\gamma_i$& $A_{1g}$& 0.785& 0.668 & 0.376   \\
& $E_{2g}^1$ & 0.483& 0.316 & 0.246 \\
& $A_{2u}^2$ & 0.494& 0.398 & 0.256 \\
\end{tabular}
\end{ruledtabular}

\end{table}

%\end{document}

\newpage
\begin{figure}
\includegraphics{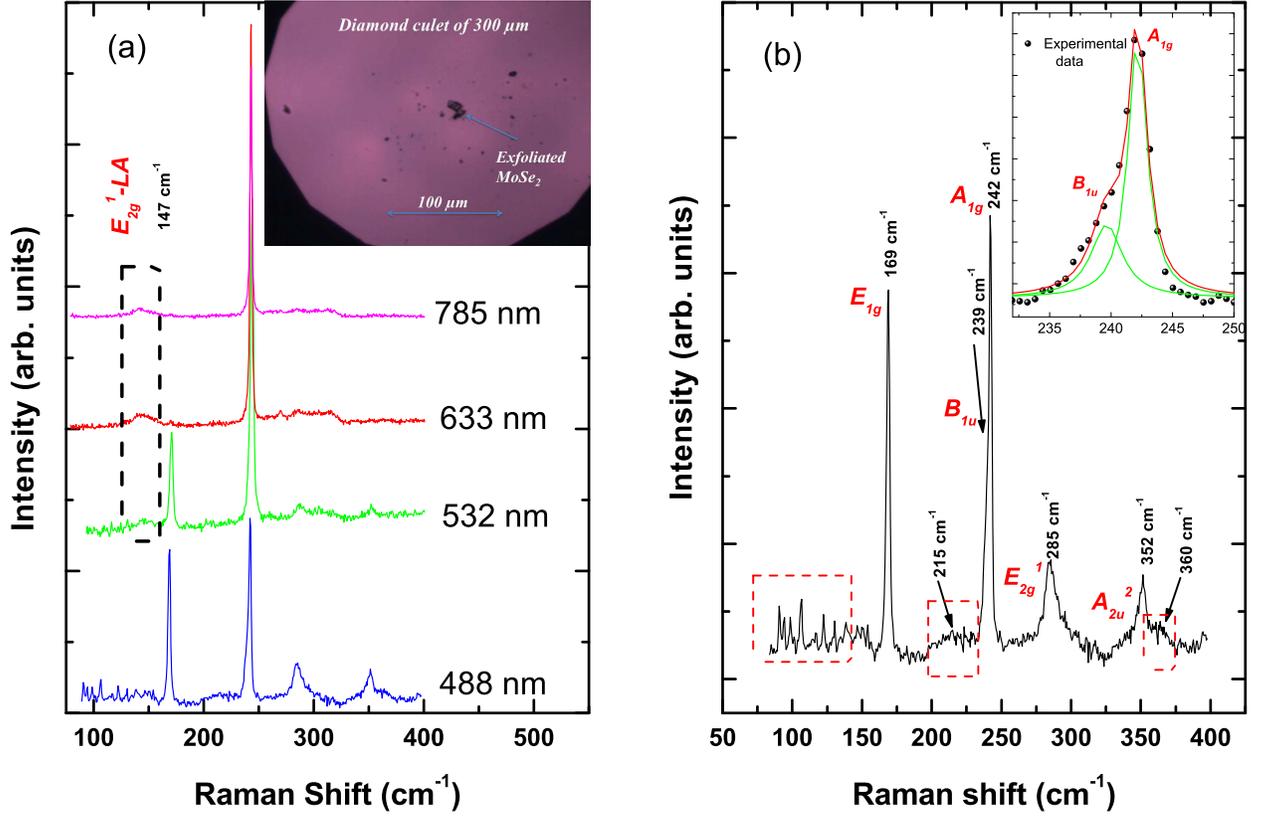}
\caption{\label{fig1}(Colour online) (a)  Room temperature Raman spectra collected in back scattering geometry of exfoliated few layered $MoSe_2$ placed on diamond anvil culet of 300 $\mu$m (magnified image shown in right-inset), using four different laser excitations. (b) Resonance Raman spectra taken using 488 nm laser source. Splitting of $A_{1g}$ mode into two parts indicating the few-layered nature of the sample is shown in right-inset. Apart from the two major Raman modes ($A_{1g}$ and $E_{2g}^1$) of $MoSe_2$, several prominent resonance Raman modes are clearly seen at 147, 169, 239, and 352 cm$^{-1}$ corresponding to $E_{2g}^1$ - $LA$, $E_{1g}$, $B_{1u}$, and $A_{2u}^2$ respectively. Few additional modes in the range from 75 to 140 $cm^{-1}$ and humps around 215 and 360 $cm^{-1}$ are indicated inside dotted red rectangle.} 

\end{figure}  

\begin{figure}
\includegraphics{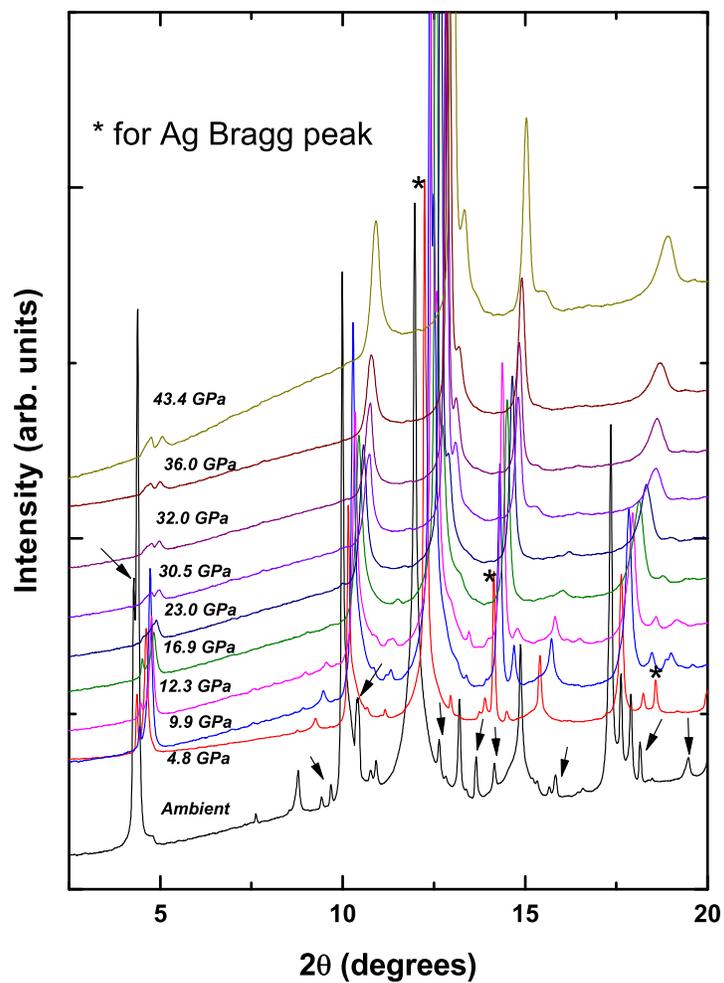}
\caption{\label{fig2}(Colour online) Pressure evolution of $XRD$ patterns of few layered $MoSe_2$ at selected pressure points. Arrows in the ambient pattern shows presence of new diffraction lines of triclinic structure. The star marks represent the silver pressure markers. Ambient XRD pattern is taken without adding silver.}
\end{figure}  

\begin{figure}
\includegraphics{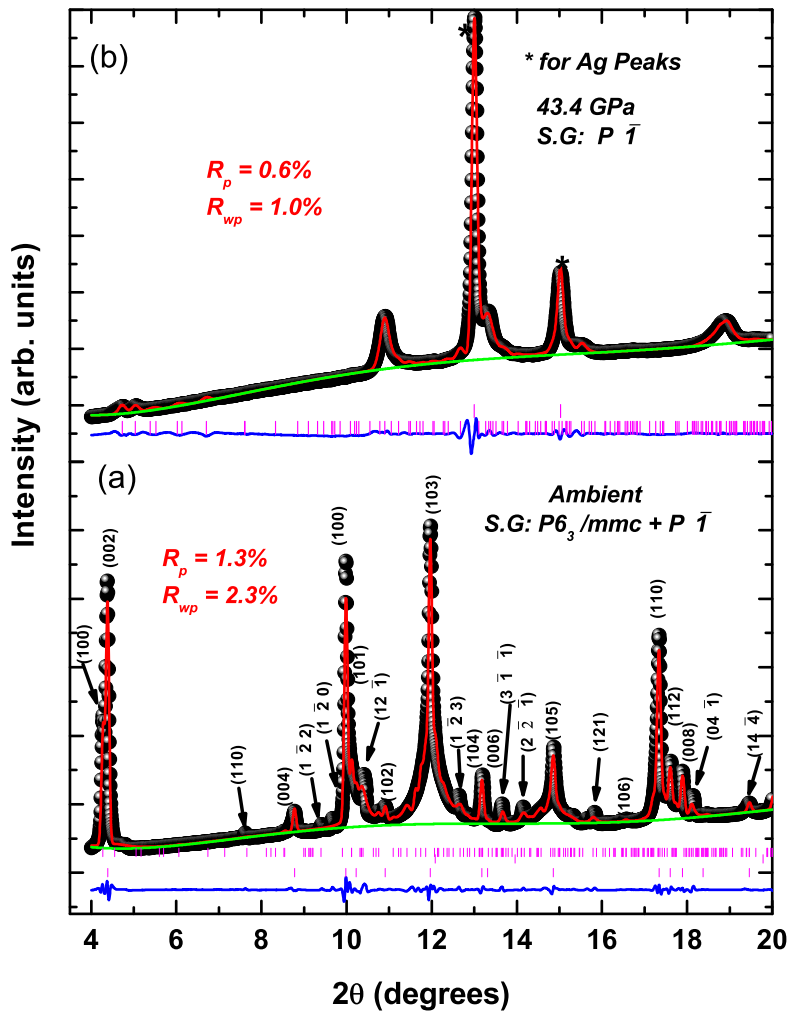}
\caption{\label{fig3}(Colour online) Lebail fit of synchrotron $XRD$ patterns of exfoliated few layered $MoSe_2$ (a) in mixed phases of hexagonal and triclinic structures  at ambient conditions, and (b) triclinic phase at 43.4 GPa. Black filled circles represent the observed data points, red line shows the fit to the data, background is shown by the green line, upper vertical ticks indicates $Ag$ Bragg peaks, with the Bragg peaks of samples marked by the ticks at a slightly lower level, and difference is marked by the blue line. The star marks represent the silver pressure markers. Ambient XRD pattern is taken without adding silver.}
\end{figure} 

\begin{figure}
\includegraphics{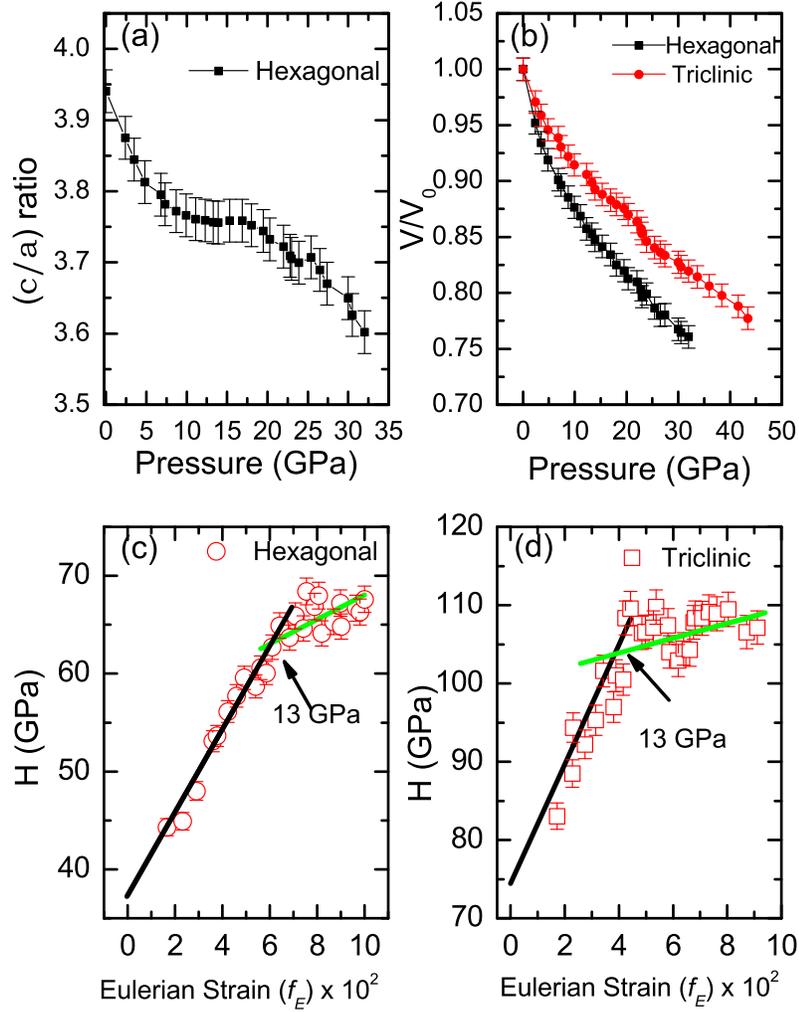}
\caption{\label{fig4}(Colour online) (a) The evolution of $c/a$-ratio with pressure for the hexagonal phase of exfoliated $MoSe_2$. (b) Pressure dependence of relative volume of both the hexagonal and triclinic phases of exfoliated $MoSe_2$ with respect to the ambient pressure value. (c) Plot of reduced pressure ($H$) versus Eulerian strain ($f_E$)  in the hexagonal phase of exfoliated $MoSe_2$, showing discontinuous change in their linear behavior around 13 GPa. (d) triclinic phase of exfoliated $MoSe_2$, showing discontinuous change in their linear behavior around 13 GPa. }
\end{figure}

\begin{figure}
\includegraphics{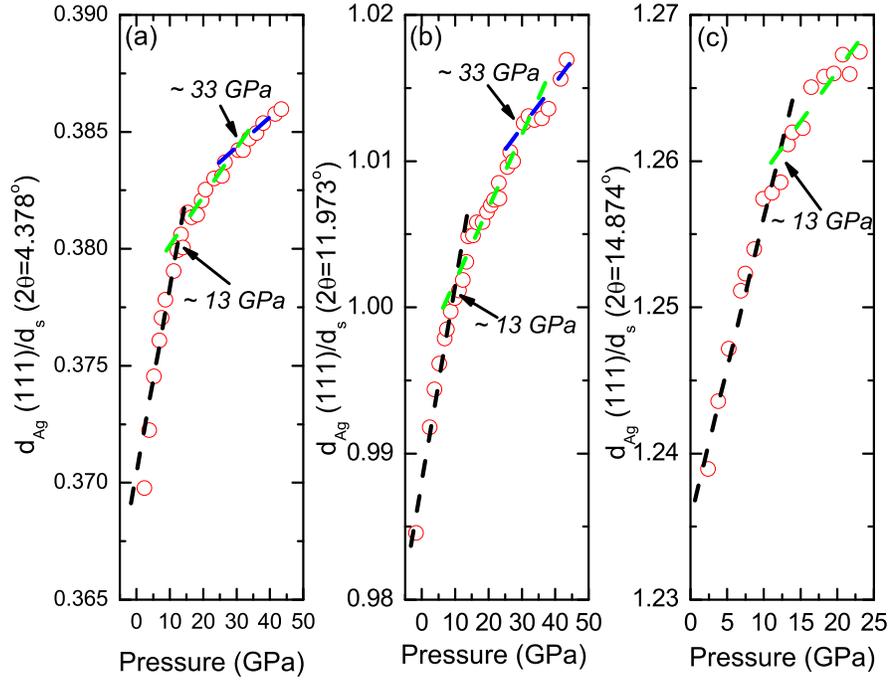}
\caption{\label{fig5}(Colour online) a) Pressure behavior of $d$-spacing corresponding to $2\Theta$ value of 4.38$^o$ with respect to the $d_{Ag}$-spacing of (111) lattice plane of Silver. The $d$-spacing corresponding to this $2\Theta$ value relates to (002) lattice plane for hexagonal phase at ambient conditions and (01$\overline{1}$) lattice plane for triclinic phase at high pressures. (b) Pressure behavior of $d$-spacing corresponding to $2\Theta$ value of 11.97$^o$ with respect to the $d_{Ag}$-spacing of (111) lattice plane of Silver. The $d$-spacing corresponding to this $2\Theta$ value relates to (103) lattice plane for hexagonal phase at ambient conditions and (02$\overline{3}$) lattice plane for triclinic phase at high pressures. (c) The pressure behavior of $d$-spacing corresponding to $2\Theta$ value of 14.87$^o$ with respect to the $d_{Ag}$-spacing of (111) lattice plane of Silver. The $d$-spacing corresponding to this $2\Theta$ value relates to (105) lattice plane for hexagonal phase at ambient conditions and (2$\overline{3}$0) lattice plane for triclinic phase at high pressures.}

\end{figure}

\begin{figure}
\includegraphics{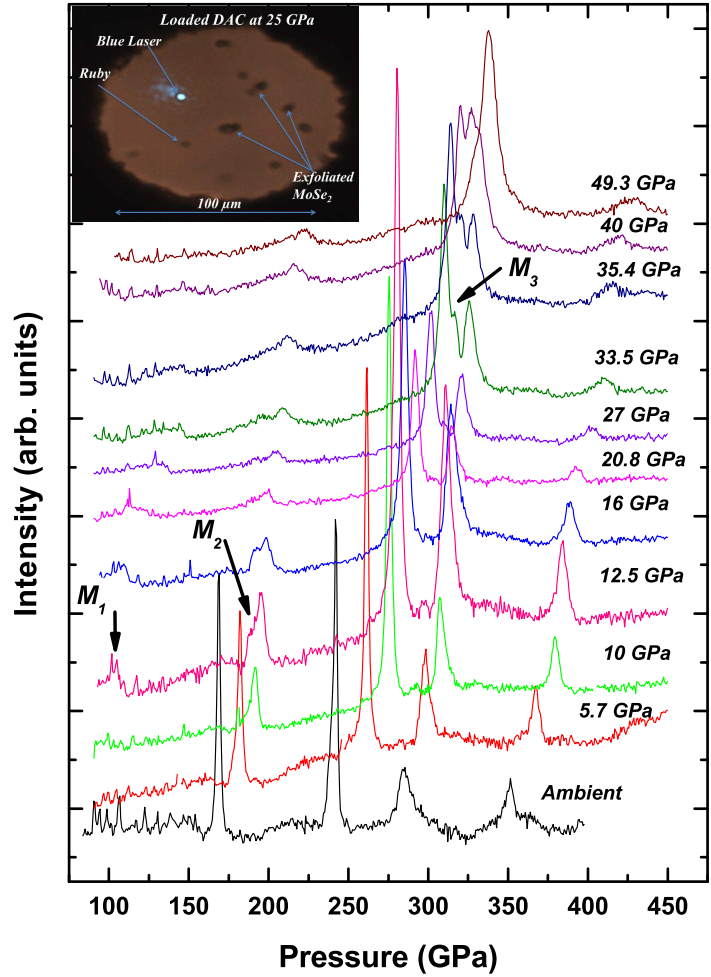}
\caption{\label{fig6}(Colour online) Raman spectra of few layered $MoSe_2$ at selected pressure points. Emergence of the mode $M_1$ and $M_2$ at 103 and 190 cm$^{-1}$ is shown by black arrows at around 13 GPa. Emergence of the another mode $M_3$ at 317 cm$^{-1}$ is also shown by black arrow at around 33 GPa. Rapid suppression of the intensity of the $A_{1g}$ mode with respect to $E_{2g}^1$ is observed above 35 GPa.  A representative picture of the loaded $DAC$ at 25 GPa for Raman measurement is shown in inset.}
\end{figure}

\begin{figure}
\includegraphics{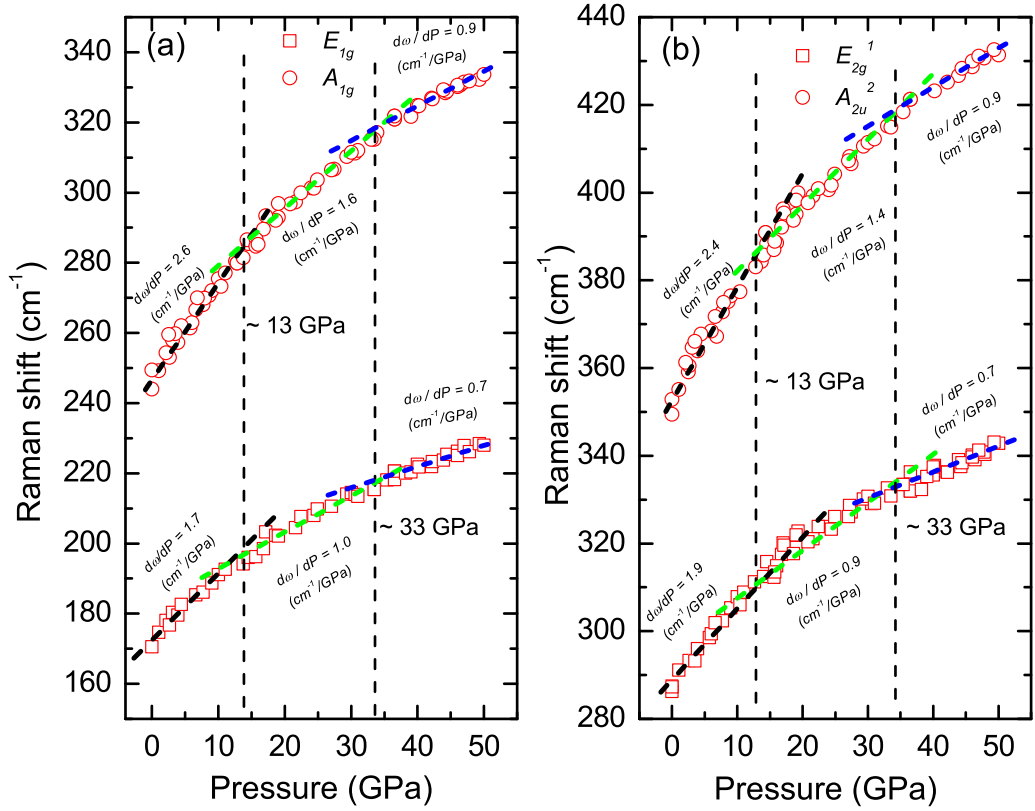}
\caption{\label{fig3}(Colour online) Linear pressure behaviour of Raman modes, (a)$A_{1g}$ and $E_{1g}$; (b) $E_{2g}^1$ and $A_{2u}^2$. Two slope changes in linear pressure behavior of all these modes are observed at around 13 and 33 GPa as indicated by dashed black line.}
\end{figure}

\begin{figure}
\includegraphics{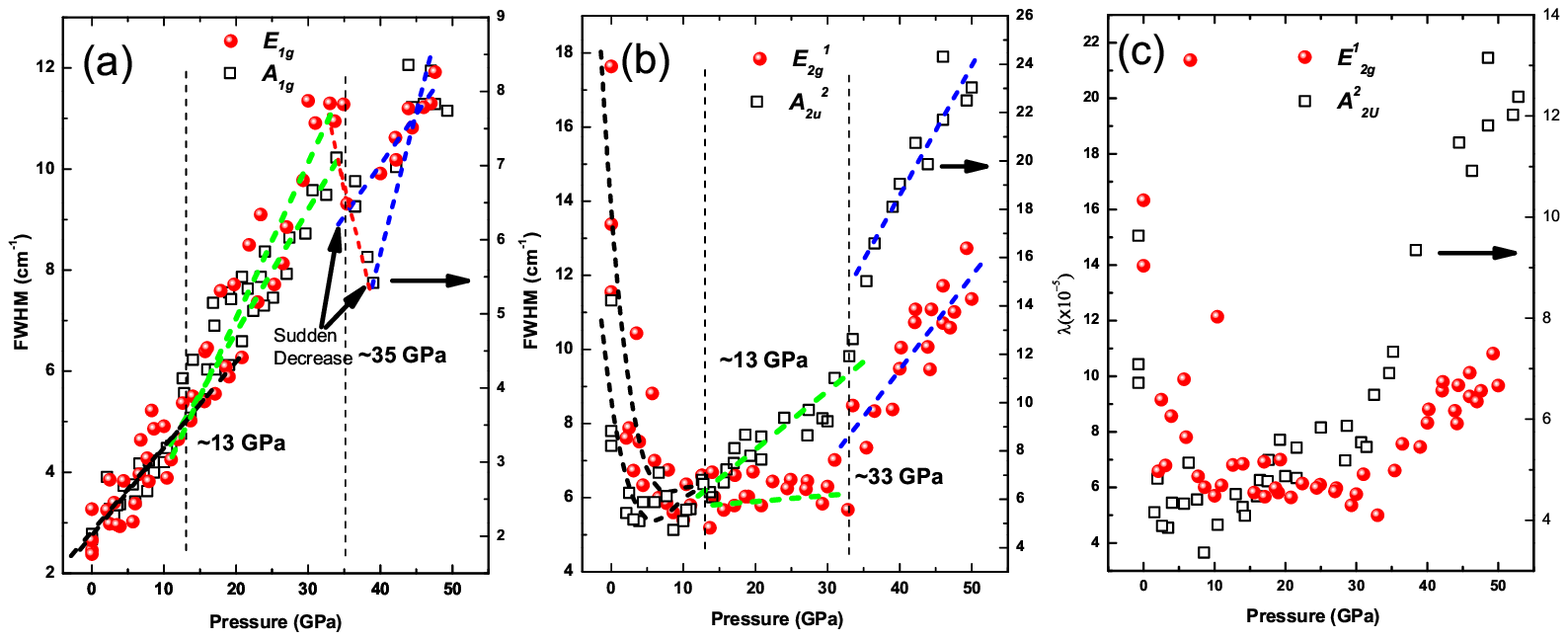}
\caption{\label{fig8}(Colour online) Pressure evolution of $FWHM$ of $A_{1g}$, $E_{1g}$, $E_{2g}^1$, and $A_{2u}^2$ Raman modes are shown in (a) and (b), respectively. $FWHM$ of $A_{1g}$ and $E_{1g}$ show a slope change around 13 GPa followed by a minimum at about 35 GPa. $FWHM$ of $E_{2g}^1$ and $A_{2u}^2$ soften with pressure and reach a minimum value around 13 GPa followed by a linear behavior and sudden jump around 33 GPa. Pressure behaviour of electron phonon coupling for $E_{1g}$  and $A^2_{2u}$ Raman modes are shown in Fig. 8(c)}
\end{figure}


\begin{thebibliography}{}
\bibitem{samanta14}
A. Samanta, T. Pandey, and A. K. Singh, Phys. Rev. B. {\bf90}, 174301 (2014).

\bibitem{huang10}
M. Huang, H. Yan, T. F. Heinz, and J. Hone, Nano Lett. {\bf10}, 4074–4079 (2010).

\bibitem{song18}
C. Song, F. Fan, N. Xuan, S. Huang, G. Zhang, C. Wang, Z. Sun, H. Wu, and H. Yan, ACS Appl. Mater. Interfaces {\bf10}, 3994−4000 (2018).

\bibitem{sharma14}
M. Sharma, A. Kumar, P. K. Ahluwalia, and R. Pandey, J. Appl. Phys. {\bf116}, 063711 (2014).

\bibitem{liu14}
Z. Liu, M. Amani, S. Najmaei, Q. Xu, X. Zou, W. Zhou, T. Yu, C. Qiu, A. G. Birdwell, F. J. Crowne, R. Vajtai, B. I. Yakobson, Z. Xia, M. Dubey, P. M. Ajayan, and J. Lou, Nat. Commun. {\bf5}, 5246 (2014).

\bibitem{duerloo14}
K.~A. N. Duerloo, Y. Li, E. J. Reed, Nat. Commun. {\bf5}, 4214 (2014).

\bibitem{nayak14}
A. P. Nayak, T. Pandey, D. Voiry, J. Liu, S. T. Moran, A. Sharma, C. Tan, C. H. Chen, L. J. Li, M. Chhowalla, J. F. Lin, A. K. Singh, and D. Akinwande, Nano Lett. {\bf15}, 346 (2014).

\bibitem{nayak2014} 
 A. P. Nayak, S. Bhattacharyya, J. Zhu, J. Liu, X. Wu, T. Pandey, C. Jin, A. K. Singh, D. Akinwande, and J. F. Lin, Nat. Commun. {\bf5}, 3731 (2014).

\bibitem{chi14}
Z. H. Chi, X. M. Zhao, H. Zhang, A. F. Goncharov, S. S. Lobanov, T. Kagayama, M. Sakata, and X. J. Chen, Phys. Rev. Lett. {\bf113}, 036802 (2014).

\bibitem{hromadova13}
L. Hromadova, R. Martonak, and E. Tosatti, Phys. Rev. B {\bf87}, 144105 (2013).


\bibitem{zhao15}
Z Zhao, H Zhang, H. Yuan, S. Wang, Y. Lin, Q. Zeng, G. Xu, Z. Liu, G.K. Solanki, K.D. Patel, Y. Cui, H. Y. Hwang, and W. L. Mao, Nat. Commun. {\bf6}, 7312 (2015).


\bibitem{riflikova14}
M. Riflikova, R. Martonak, and E. Tosatti, Phys. Rev. B {\bf 90}, 035108 (2014).

\bibitem{bandaru14}
N. Bandaru, R. S. Kumar, J. Baker, O. Tschauner, T. Hartmann, Y. Zhao, and R. Venkat, Int. J. Mod. Phys. B {\bf28}, 1450168 (2014).

\bibitem{duwal16}
S. Duwal, and C. S. Yoo, J. Phys. Chem. C. {\bf120}, 5101 (2016).

\bibitem{nayak15} 
 A. P. Nayak, Z. Yuan, B. Cao, J. Liu, J. Wu, S. T. Moran, T. Li, D. Akinwande, C. Jin, and J. F. Lin, ACS Nano {\bf9}, 9117 (2015).


\bibitem{saha18}
P. Saha, B. Ghosh, R. Jana, and G. D. Mukherjee, J. Appl. Phys. {\bf 123}, 204306 (2018).


\bibitem{sugai82}
S. Sugai, and T. Ueda, Phys. Rev. B {\bf26}, 6554 (1982).

\bibitem{dave04}
M. Dave, R. Vaidya, S. G. Patel, and A. R. Jani, Bull. Mater. Sci. {\bf 27}, 213 (2004).

\bibitem{aksoy08}
R. Aksoy, E. Selvi, and Y. Ma, J. Phys. Chem. Solids {\bf 69}, 2138 (2008).


\bibitem{kohulak17}
O. Kohulak, and R. Martonak, Phys. Rev. B {\bf 95}, 054105 (2017).


\bibitem{pal17}
B. Pal, A. Singh, Sharada G., P. Mahale, A. Kumar, S. Thirupathaiah, H. Sezen, M. Amati, L. Gregoratti, U. V. Waghmare, and D. D. Sarma, Phys. Rev. B {\bf96}, 195426 (2017).

\bibitem{mayoral16}
V. V. Mayoral, C. Backes, D. Hanlon, U. Khan, Z. Gholamvand, M. O'Brien, G. S. Duesberg, C. Gadermaier, and J. N. Coleman, Adv. Funct. Mater, {\bf26}, 1028 (2016).

\bibitem{mao86}
H. K. Mao, J. Xu, and P. M. Bell, J. Geophys. Res. {\bf91}, 4673 (1986). 

\bibitem {dewaele08}
A. Dewaele, M. Torrent, P. Loubeyre, and M. Mezouar, Phys. Rev. B {\bf78}, 104102 (2008).

\bibitem{prescher15}
C. Prescher, and V. B. Prakapenka, High Press. Res. {\bf35}, 223 (2015).

\bibitem{shirley02}
R. Shirley, The CRYSFIRE 2002 System for Automatic Powder Indexing: Users Manual (TheLattice Press, Guildford, 2002).

\bibitem{checkcell}
 LMGP-Suite Suite of Programs for the interpretation of X-ray Experiments, by Jean laugier and Bernard Bochu, ENSP/Laboratoire des Matériaux et du Génie Physique, BP 46. 38042 Saint Martin d'Hères, France.

\bibitem{toby01}
 B. H. Toby, EXPGUI, J. Appl. Crystallogr. {\bf34}, 210 (2001).

\bibitem{sekine80}
T. Sekine, M. Izumi, T. Nakashizu, K. Uchinokura, and E. Matssura, J. Phys. Soc. Jpn. {\bf49}, 1069 (1980).

\bibitem{nam15}
D. Nam, J. U. Lee, and H. Cheong, Sci. Rep. {\bf5}, 17113 (2015).

\bibitem{soubelet16}
P. Soubelet, A. E. Bruchhausen, A. Fainstein, K. Nogajewski, and C. Faugeras, Phys. Rev. B {\bf93}, 155407 (2016).


\bibitem{larentis12}
s. Larentis, B. Fallahazad, and E. Tutuc, Appl. Phys. Lett. {\bf101}, 223104 (2012).

\bibitem{cheong00}
H. M. Cheong, Y. Zhang, A. Mascarenhas, and J. F. Geisz, Phys. Rev. B {\bf61}, 13687 (2000).

\bibitem{mahler14}
B. Mahler, V. Hoepfner, K. Liao, and G. A. Ozin, J. Am. Chem. Soc. {\bf136}, 14121 (2014).

\bibitem{angel01}
R. J. Angel,  High-pressure, High-Temperature Crystal Chemistry, Rev. Mineral. Geochem.  Edited by R. M. Hazen and R. T. Downs, CRC Press, Boka Raton, Fla. {\bf41},  pp. 35 , (2001).

\bibitem{poilan11}
A. Polian, M. Gauthier, S. M. Souza, D. M. Triches, J. Cardoso de Lima, T. A. Grandi, Phys. Rev. B {\bf83} 113106 (2011).


\bibitem{birch47}
F. Birch, Phys. Rev. B {\bf71}, 809 {1947}.

\bibitem{murnaghan37}
F. D. murnaghan, American Journal of Mathematics {\bf59}, 235 (1937).

\bibitem{jana16}
R. Jana, P. Saha, V. Pareek, A. Basu, S. Kapri, S. Bhattacharyya, and  G. D. Mukherjee, Sci. Rep. {\bf6}, 31610 (2016).

\bibitem{ardit10}
M. Ardit, G. Cruciani, M. Dondi, M. Merlini, and P. Bouvier, Phys. Rev. B {\bf82}, 064109 (2010).


\bibitem{caramazza17}
S. Caramazza, F. Capitani, C. Marini, L. Malavasi, P. Dore, P. Postorino, IOP Conf. Series: J. Phys. Conf. Series {\bf950}, 042012 (2017).



\bibitem{bera13}
A. Bera, K. Pal, D. V. S. Muthu, S. Sen, P. Guptasarma, U. V. Waghmare, and A. K. Sood, Phys. Rev. Lett. {\bf110}, 107401 (2013).

\bibitem{gupta17}
S. N. Gupta, A. Singh, K. Pal, B. Chakraborti, D. V. S. Muthu, U. V. Waghmare, and A. K. Sood, Phys. Rev. B. {\bf96}, 094104 (2017).











\end{thebibliography}
\end{document}